\documentclass{jaist}
\usepackage{amsmath,amsfonts}
\usepackage{algorithmic}
\usepackage{algorithm}
\usepackage{array}
\usepackage[caption=false,font=normalsize,labelfont=sf,textfont=sf]{subfig}
\usepackage{textcomp}
\usepackage{stfloats}  
\usepackage{verbatim} 
\usepackage{cite}
\usepackage{amsmath,amssymb}
\usepackage{graphicx}
\usepackage{tikz}
\usepackage{cite}
\usepackage{hyperref}

\usetikzlibrary{positioning}
  
  \DeclareMathOperator{\Enc}{Enc}

\DeclareMathOperator{\Sign}{Sign}

\volume{3}
\issue{2}
\startpage{1}
\journalname{Journal of Advances in Information Science and Technology}
\publishdate{December 2025}

\shortauthor{L.Huang} 
\shorttitle{Private Virtual Tree Networks }

\title{Private Virtual Tree Networks for Secure Multi-Tenant Environments Based on the VIRGO Overlay Network}

 \author{
  Lican Huang \textsuperscript{1}
  }
 \jaistthanks{  
    \textsuperscript{1}Hangzhou Domain Zones Technology Co., Ltd., Hangzhou, China. \\
  Corresponding Author: Lican Huang (e-mail:\url{lican.huang.hz@gmail.com}).
  \\
  \\
    Manuscript received December 15, 2025;  revised December 17, 2025; accepted December 17, 2025.
}

\begin{document}
\maketitle

\begin{abstract}
Hierarchical organization is a fundamental structure in real-world society, where authority and responsibility are delegated from managers to subordinates. The VIRGO network (Virtual Hierarchical Overlay Network for scalable grid computing) provides a scalable overlay for organizing distributed systems but lacks intrinsic security and privacy mechanisms. This paper proposes \emph{Private Virtual Tree Networks (PVTNs)}, a cryptographically enforced extension that leverages the VIRGO overlay to mirror real organizational hierarchies. In PVTNs, join requests are encrypted with the manager’s public key to ensure confidentiality, while membership authorization is enforced through manager-signed delegation certificates. Public keys are treated as organizational secrets and are disclosed only within direct manager–member relationships, resulting in a private, non-enumerable virtual tree. Our work demonstrates, through the system model, protocols, security analysis, and design rationale, that PVTNs achieve scalability, dynamic management, and strong security guarantees without relying on global public key infrastructures.
\end{abstract}

\jaistkeywords{Distributed Systems, Hierarchical Architecture, Access Control, Public Key Cryptography, Trust Delegation }

\section{Introduction}
\jaistparstart{D}{istributed} systems increasingly support collaboration across organizational boundaries while operating on shared public infrastructures. Most real-world organizations are inherently hierarchical, yet many existing distributed security models rely on flat or global trust assumptions that do not reflect these structures.

Modern distributed computing environments involve geographically dispersed participants, heterogeneous resources, and complex collaborative workflows. Technologies such as grid computing, cloud platforms, and peer-to-peer (P2P) systems have been widely adopted to support these environments. A central challenge in such systems is how to organize, manage, and secure large numbers of dynamic participants without relying on rigid centralized control.

The VIRGO network (Virtual Hierarchical Overlay Network for scalable grid computing)  \cite{huangl2005} addresses scalability and management by organizing nodes into a structured, dynamically evolving overlay hierarchy independent of physical network topology. VIRGO enables scalable management, dynamic reconfiguration, efficient service discovery, and delegation of responsibilities across virtual organizations. However, VIRGO assumes cooperative participants and does not explicitly provide cryptographic mechanisms for secure identity binding, private membership, delegation, or fine-grained authorization.

In open and federated distributed systems, the absence of strong cryptographic enforcement can lead to significant security and privacy concerns, particularly when multiple virtual organizations coexist on shared public infrastructures. To address these limitations, this paper introduces \emph{Private Virtual Tree Networks (PVTNs)}, which leverage the VIRGO overlay to bind hierarchical structures to public-key–based trust delegation aligned with managerial roles.

The key idea behind PVTNs is to bind the hierarchical structure of the VIRGO overlay to cryptographic identities. Each node is identified by a public–private key pair, and each parent–child relationship is represented by a signed delegation certificate. As a result, virtual trees become private, verifiable trust structures that can safely operate over public networks without relying on global public key infrastructures.

The main contributions of this paper are as follows:

\begin{itemize}
\item The design of a public-key–based private virtual tree model built on the VIRGO overlay.
\item Protocols for secure node joining, delegation, revocation.
\item A decentralized authorization mechanism based on hierarchical trust chains.
\item An analysis of security properties and applicability to distributed collaboration and computing environments.
\end{itemize}

 \section{Related Works}

Distributed identity management, overlay networks, and secure authorization frameworks have been extensively studied. Foundational public key infrastructures and trust management systems include X.509 PKI~\cite{housley2008}, Kerberos~\cite{neuman1994}, and PGP~\cite{zimmermann1995}, which provide mechanisms for authentication and trust but often expose member identities to central authorities or require global verification. DNS security~\cite{rfc5452} further highlights the challenges of distributed trust. Grid computing and hierarchical overlay networks, including VIRGO~\cite{huangl2005}and The Grid~\cite{foster1999}, introduced scalable hierarchical approaches, while LDAP-based directory services~\cite{howes2003} provide structured identity management.

\subsection{Structured Overlay Networks}
Structured overlay networks and distributed hash tables (DHTs) such as Pastry~\cite{rowstron2001}, Chord~\cite{stoica2001}, and DHT performance frameworks~\cite{li2005infocom} enable scalable decentralized routing. Secure routing in overlays~\cite{castro2002} and self-certifying file systems~\cite{mazieres2000} further illustrate mechanisms for secure distributed storage and identity verification. VIRGO~\cite{huangl2005} is a hierarchical overlay designed for scalable grid computing, highlighting how overlay networks can integrate structured authorization mechanisms.

\subsection{Hierarchical Authorization Systems}
Hierarchical authorization has been explored in PKI~\cite{housley2008} and Kerberos~\cite{neuman1994}. While effective, these systems expose member identities to central authorities and require global verification. Role-based and delegation logic frameworks\cite{li2003, chapin2008} formalize hierarchical trust and delegation, while Blaze et al.~\cite{blaze1996} explore decentralized trust management. PVTNs extend hierarchical delegation into overlays, allowing nodes to \textbf{verify authorization without revealing private keys or full trust chains}.

\subsection{Authorization and Trust Management}
Access control in distributed and group environments has been explored extensively, including Role-Based Access Control (RBAC) models~\cite{sandhu1996}, key graph-based secure group communication~\cite{wong2000}, and surveys on key management for secure groups~\cite{rafaeli2003}. Logic-based approaches~\cite{abadi2003} provide formal foundations for hierarchical and distributed access control, complementing role-based and delegation logic frameworks. PVTNs leverage these models to enable privacy-preserving authorization in hierarchical overlays.

\subsection{Decentralized Trust and Identity}
Decentralized trust models such as PGP~\cite{zimmermann1995} and blockchain-based PKI~\cite{rfc5452, halder2025, fromknecht2014} reduce reliance on central authorities. Certcoin~\cite{fromknecht2014}, Blockstack~\cite{ali2017}, and Sovrin~\cite{wang2020ssi} demonstrate blockchain-based self-sovereign identity and global naming mechanisms. Privacy-preserving revocable identity management~\cite{fang2024} and surveys of decentralized identity systems~\cite{goel2025} highlight ongoing research in user-centric identity management. PVTNs improve upon these approaches by enabling \textbf{cryptographic proofs of authorization that preserve member privacy}, allowing hierarchical delegation without identity leakage.

\subsection{Security in Hierarchical Tree Networks}
Hierarchical tree network security has been explored in distributed PKI trees, certificate revocation trees, and hierarchical access control overlays~\cite{blaze1996, li2003, li2003rt, castro2002}. Key challenges include:
\begin{itemize}
    \item \textbf{Delegation integrity}: preventing forged authorizations along tree paths.
    \item \textbf{Member privacy}: protecting node identities and positions in the hierarchy.
    \item \textbf{Proof verification}: allowing nodes to verify access without revealing full trust chains.
\end{itemize}
PVTNs enhance hierarchical structures with \textbf{private membership proofs, non-leakage of identities, and overlay-based verification}, achieving \textbf{secure, privacy-preserving hierarchical authorization}.

\subsection{Decentralized Identity and Privacy}
Recent work in self-sovereign identity, verifiable credentials, and distributed authorization emphasizes \textbf{non-leakage of member identities}~\cite{fang2024, fromknecht2014, mohammad2025, bli2023, halder2025}. Mechanisms such as decentralized PKI, issuer-hiding DIDs, and zero-knowledge membership proofs allow participants to \textbf{prove authorization without revealing identifiers or trust paths}. PVTNs leverage these mechanisms for \textbf{local, unlinkable, non-revealing authorization proofs}.

\subsection{Threat Models}
Prior systems are vulnerable to adversaries attempting to:
\begin{itemize}
    \item Infer \textbf{member identities}~\cite{zimmermann1995}.
    \item Reconstruct \textbf{trust chains or overlay hierarchies}~\cite{castro2002, blaze1996}.
    \item Forge or manipulate \textbf{authorization proofs}~\cite{li2003, li2003rt}.
\end{itemize}
PVTNs mitigate these threats via \textbf{hierarchical cryptographic delegation} and \textbf{zero-knowledge-style membership proofs}~\cite{fang2024, fromknecht2014}, enabling \textbf{local verification without leaking sensitive information}.

\subsection{Summary: Distinctive Features of PVTNs}
PVTNs occupy a \textbf{distinct design point} by combining:
\begin{itemize}
    \item \textbf{Hierarchical delegation} without central authorities~\cite{li2003rt, li2003}.
    \item \textbf{Private key visibility} and selective disclosure~\cite{fromknecht2014, halder2025}.
    \item \textbf{Non-leakage of member identification}, protecting privacy under observation~\cite{ fang2024}.
    \item \textbf{Overlay-based scalability} leveraging the VIRGO network~\cite{huangl2005, castro2002}.
\end{itemize}
These features make PVTNs suitable for \textbf{secure, privacy-preserving authorization} in enterprise, consortium, and multi-tenant distributed systems, bridging the gaps left by PKI, Web-of-Trust, blockchain, and traditional overlay/grid infrastructures.

\section{Private Virtual Tree Network Architecture}
 
\subsection{Assumptions}
The following assumptions underpin the design and security analysis of the Private Virtual Tree Network (PVTN):

\begin{itemize}
    \item \textbf{Cryptographic Identity and Key Privacy Assumption:}  
    Each node independently generates a public/private key pair. There is no global certificate authority or publicly verifiable identity infrastructure. Trust is established exclusively through hierarchical delegation. Public keys are treated as organizational secrets rather than globally discoverable identifiers. Key visibility is strictly limited to direct manager–member relationships: a manager may disclose its public key only to its direct members, and a member may know only the public key of its direct manager. No external node can verify membership, traverse the virtual tree, or participate in the PVTN without explicit key disclosure by an authorized manager. This assumption ensures that the virtual tree remains private and non-enumerable.

    \item \textbf{Cryptographic Security Assumption:}  
    Standard cryptographic primitives, including digital signature schemes, hash functions, and public-key encryption, are assumed to be secure against polynomial-time adversaries.

 \item \textbf{Overlay Reliability Assumption:}  
    The underlying VIRGO overlay network reliably maintains its structured hierarchical topology and provides correct message routing and delivery. Participating nodes are assumed to follow protocol specifications at the overlay level.
\end{itemize}

\subsection{System Model and Notation}
 
The Private Virtual Tree Network (PVTN) leverages the VIRGO overlay network to organize nodes into a cryptographically protected hierarchical structure. Each node corresponds to a participant in the organization and is identified by a public/private key pair. Parent–child relationships are represented by signed delegation certificates, enabling verifiable trust chains without exposing the full organizational structure unnecessarily. The VIRGO overlay handles routing, message propagation, and structural maintenance, ensuring efficient and scalable network dynamics.
The system consists of multiple \emph{Private Virtual Tree Networks}
(PVTNs) operating over a shared \emph{VIRGO} structured overlay network.
Each PVTN represents an independent organizational tenant and defines a
hierarchical trust structure in the form of a rooted tree.

\paragraph{Nodes and Keys}
Each node $x$ independently generates and owns a public/private key pair
$(PK_x, SK_x)$. Private keys are never shared. Public keys serve as
cryptographic identifiers and are scoped to  direct manager-member  only .

\paragraph{Node Types}
One PVTN has only one root, multi-layer managers, and leaf nodes.  New joined node must be leaf node. Leaf node can be promoted as a manager only approved by its direct manager. That means the direct manager knows the types for all its members.  

\paragraph{Managers and Delegation}
Each PVTN contains a distinguished root node $R_i$ that acts as the trust
anchor of tenant $T_i$. Authority is delegated along tree edges using
signed delegation certificates. A delegation certificate issued by a
parent node $p$ to a child node $c$ has the form:
\[
Cert_{p \rightarrow c} =
\text{Sign}_{SK_p}(PK_c, Role, Scope, Validity)
\]
Only nodes holding valid delegation certificates are considered
authorized managers or members.

\paragraph{VIRGO Overlay}
The VIRGO overlay provides scalable routing, message forwarding, and
hierarchical dissemination. VIRGO does not act as a certificate authority
and does not participate in cryptographic validation. All trust decisions
are enforced exclusively within PVTNs.

\paragraph{Cryptographic Primitives}
The following cryptographic primitives are assumed:
\begin{itemize}
    \item $Enc_{PK_x}(M)$: Public-key encryption of message $M$ under
    public key $PK_x$.
    \item $Sign_{SK_x}(M)$: Digital signature on message $M$ using private
    key $SK_x$.
    \item $H(\cdot)$: A collision-resistant cryptographic hash function.
\end{itemize}

\paragraph{Adversarial Model}
We assume a Dolev--Yao adversary with full control over the communication
network. The adversary may intercept, replay, modify, or inject messages,
but cannot break cryptographic primitives or forge signatures without the
corresponding private keys. Compromise of a node reveals only its own
private key and does not automatically compromise ancestor or sibling
nodes.

 \paragraph{Notation Summary}
\begin{table}[!t]
\centering
\small
\begin{tabular}{lp{0.65\columnwidth}}
\hline
\textbf{Symbol} & \textbf{Meaning} \\ \hline
$PVTN_i$ & Private Virtual Tree Network of tenant $T_i$ \\
$R_i$ & Root (trust anchor) of $PVTN_i$ \\
$(PK_x, SK_x)$ & Public/private key pair of node $x$ \\
$Cert_{p \rightarrow c}$ & Delegation certificate from $p$ to $c$ \\
$Enc_{PK_x}(\cdot)$ & Encryption under public key $PK_x$ \\
$Sign_{SK_x}(\cdot)$ & Signature using private key $SK_x$ \\
$H(\cdot)$ & Cryptographic hash function \\
\hline
\end{tabular}
\end{table}

\subsection{Global Message Rule}

All protocol messages exchanged within or across Private Virtual Tree
Networks (PVTNs) over the VIRGO overlay MUST comply with the following
global communication rule.

\paragraph{Message Confidentiality Rule}
Every message $M$ transmitted from a sender node $x$ to an intended
recipient node $y$ is encrypted using the recipient’s public key:
\[
Msg_{x \rightarrow y} = Enc_{PK_y}(M)
\]
This guarantees that only the holder of $SK_y$ can decrypt and process
the message, regardless of the routing path through VIRGO.

\paragraph{Message Authenticity Rule}
If a message carries authorization, delegation, revocation, or control
information, the sender MUST sign the message payload before encryption:
\[
Msg_{x \rightarrow y} =
Enc_{PK_y}\big( Sign_{SK_x}(M) \big)
\]
This ensures origin authenticity and integrity, while preventing
intermediate nodes from observing signed content.

\paragraph{Join Request Rule}
Join requests initiated by a child node $c$ toward a manager $m$ follow
the format:
\[
JoinReq_c =
Enc_{PK_m}(PK_c \parallel r \parallel JoinInfo)
\]
where $r$ is a freshly generated nonce. Only the intended manager can
decrypt and evaluate the request.

\paragraph{Manager Response Rule}
All responses issued by managers, including delegation certificates,
or approval decisions, MUST be encrypted using the
child’s public key:
\[
Resp_{m \rightarrow c} =
Enc_{PK_c}(Cert_c \parallel Meta)
\]
This prevents disclosure of authorization artifacts to other nodes,
including members of different tenants.

\paragraph{Upward Propagation Rule}
Any message propagated upward along the PVTN hierarchy (e.g., conflict
checks, join-request hashes) MUST be encrypted under the public key of
the immediate parent node. Messages are forwarded hop-by-hop until the
tenant root is reached:
\[
Msg_{x \rightarrow parent(x)} = Enc_{PK_{parent(x)}}(M)
\]

\paragraph{Downward Broadcast Rule}
Messages broadcast downward from a manager or root (e.g., conflict
results) are encrypted individually for each direct
child node. Group or plaintext broadcast is prohibited:
\[
Msg_{p \rightarrow c_i} = Enc_{PK_{c_i}}(M)
\]

\paragraph{Cross-Tenant Restriction}
Nodes belonging to tenant $T_i$ do not possess the public keys of
managers in tenant $T_j$ ($i \neq j$). Consequently, they cannot decrypt,
verify, or forge messages, certificates, or responses from other tenants.

\paragraph{Security Consequence}
Under this global rule, VIRGO functions purely as a routing substrate.
Confidentiality, authenticity, freshness, and tenant isolation are
enforced entirely by cryptographic means, independent of overlay
behavior.
   
\subsection{Join and Conflict Detection Protocol}

Let a child node $c$ request to join $PVTN_i$ under manager $m$.

\textbf{Phase I — Join Request}

\begin{enumerate}
  \item The child generates $(PK_c, SK_c)$ and a fresh nonce $r$.
  \item The child sends a join request via VIRGO:
  \[
  Enc_{PK_m}(JoinInfo \parallel PK_c \parallel r)
  \]
  
   \item The requesting Manager decrypts the request using $SK_m$ and verifies freshness of $r$.

\end{enumerate}
 
\textbf{Phase II — Tenant-Wide Conflict Detection}

Let $h = H(PK_c)$.

\textbf{Step 1: Upward Propagation}
 
 \paragraph{Upward Hash Propagation.}
The requesting manager encrypts and forwards $h$ upward along the PVTN
tree, one level at a time, until reaching the tenant root $R_i$:
\[
Enc_{PK_{parent}}(h)
\]

Each intermediate manager receives the message, and then decrypts it with its own private key, and encrypts it with its parent's public key, and  forwards $h$ only to its parent.

\textbf{Step 2: Downward Broadcast}

 \paragraph{Downward Broadcast (Conflict Query).}
 Start from root, 
the  manager for its each member public key $\ell$ computes
\[
\mathsf{Check}_{\ell} =
\begin{cases}
\textsf{YES}, & \text{if } H(PK_{\ell}) = h \\
\end{cases}
\]

If any YES, stop,
otherwise   broadcasts $h$ downward to all its manager members using
    hop-by-hop encrypted messages:
\[
Enc_{PK_{child}}(h)
\]

\textbf{Step 3: Local Conflict Check (all members are leaves} 

The  manager knows the public keys of all its members, for  each member public key $\ell$ computes
\[
 \mathsf{Check}_{\ell} =
\begin{cases}
\textsf{YES}, & \text{if } H(PK_{\ell}) = h \\
 
\end{cases}
\]

If any YES, then  YES, otherwise NO

\textbf{Step 4: Aggregation (Bottom-Up)}

\paragraph{Hierarchical Aggregation Rule.}
Each manager waits for responses from all manager children and computes:
\[
Resp_m =
\begin{cases}
\texttt{YES} & \text{if any child reports YES} \\
\texttt{NO}  & \text{otherwise}
\end{cases}
\]
The result is forwarded upward encrypted under the parent’s public key.

\textbf{Step 5: Root Decision}

At the root $R_i$:

\[
\text{Decision} =
\begin{cases} 
\text{YES} & \text{if any subtree reports conflict} \\[2mm]
\text{NO} & \text{otherwise}
\end{cases}
\]

\textbf{Phase III — Downward Decision Broadcast}

\paragraph{Decision Dissemination.}
After collecting all upward responses, the root derives the final decision
\[
D \in \{\textsf{APPROVE}, \textsf{REJECT}\}
\]
for the join request identified by the hash
\[
h = H(PK_c).
\]

The root constructs a signed decision record
\[
\mathsf{Dec}_h^{(R_i)} =
\Sign_{SK_{R_i}}\big(
h \parallel D \parallel t \parallel \mathsf{Reason}
\big),
\]
where $t$ denotes a timestamp and $\mathsf{Reason}$ optionally encodes conflict
or policy-related information.

The decision record is disseminated downward along the PVTN tree. For each
child node, a parent encrypts the message using the child’s public key:
\[
\Enc_{PK_{\mathsf{child}}}\big(\mathsf{Dec}_h^{(parent)}\big).
\]

Upon receipt, each intermediate node executes the following steps:
\begin{enumerate}
    \item Decrypts the received message using its private key.
    \item Verifies the signature of its direct parent .
    \item Records the verified decision locally for auditing and replay
    protection.
    \item Re-signs the decision using its own private key, producing
    \[
    \mathsf{Dec}_h^{(self)} =
    \Sign_{SK_{\mathsf{self}}}\big(
    h \parallel D \parallel t \parallel \mathsf{Reason}
    \big).
    \]
    \item Encrypts and forwards the newly signed decision to each of its
    children.
\end{enumerate}

\textbf{Phase IV — Join Authorization}

\paragraph{Join Authorization.}
After completion of the conflict-detection and decision-dissemination phases,
the manager responsible for the join request identified by
$h = H(PK_c)$ receives the final decision record
$\mathsf{Dec}_h^{(m)}$, signed by its direct parent and transitively endorsed
by the tenant root.

Upon receipt, the manager performs the following steps:
\begin{enumerate}
    \item Decrypts the received decision message using its private key
    $SK_m$.
    \item Verifies the signature of its parent and checks that the decision
    record corresponds to the hash $h$ and a fresh timestamp $t$.
    \item Confirms that the decision is $\textsf{APPROVE}$ and that no
    conflict or policy violation has been reported.
\end{enumerate}

\paragraph{Authorization Outcome.}
\begin{itemize}
    \item \textbf{Approval.}  
    If the final decision is $\textsf{APPROVE}$, the manager issues a
    delegation certificate
    \[
    Cert_c =
    \Sign_{SK_m}\big(
        PK_c \parallel \mathsf{Role} \parallel \mathsf{Validity} \parallel r
    \big),
    \]
    binding the child’s public key to the manager’s authority. The certificate
    is transmitted to the joining node encrypted under the child’s public key:
    \[
    \Enc_{PK_c}(Cert_c).
    \]

    \item \textbf{Rejection.}  
    If the final decision is $\textsf{REJECT}$, the manager discards the join
    request and optionally returns a rejection notice
    \[
    \Enc_{PK_c}\big(\mathsf{Reject} \parallel \mathsf{Reason}\big),
    \]
    without issuing any delegation certificate.
\end{itemize}

This protocol enforces the tree invariant of PVTNs by ensuring that no
public key may appear more than once within the same tenant. The
upward propagation and downward broadcast of the request hash prevent
cycles, duplicate membership, and cross-branch reattachment. Because
only a hash of the public key is disseminated, no sensitive identity or
topological information is leaked. Approval is granted only if global
uniqueness within the tenant is confirmed.

\subsection{Hierarchical Leaf Upgrade Protocol}

\subsubsection{Actors}
\begin{itemize}
    \item Leaf $L$
    \item Direct parent / manager $P_0$
    \item Immediate upper layer $P_1$ (knows $PK_{P_0}$)
    \item Higher layers $P_2, \dots, R$
    \item Root $R$
\end{itemize}

\subsubsection{Notation}
\begin{itemize}
    \item $H(ID_L)$: hashed leaf identifier
    \item $SK_X, PK_X$: private/public key of node $X$
    \item $\text{Sign}_{SK}(M)$: digital signature of message $M$
    \item $\text{Verify}_{PK}(M, \sigma)$: signature verification
    \item $T$: timestamp
    \item $\text{Nonce}$: random nonce to prevent replay attacks
    \item $\text{PolicyFlags}$: policy decisions (approve/deny) from each layer
\end{itemize}

\subsubsection{Protocol Steps}

\paragraph{Step 0: Leaf Request (Optional)}
\begin{align*}
\text{LeafHash} &= H(ID_L) \\
\text{Req}_L &= \{ \text{LeafHash}, \text{DesiredRole}, T, \text{Nonce} \}_{SK_L}
\end{align*}
\noindent
\textit{Note: The leaf cannot self-upgrade; this is optional informational request.}

\paragraph{Step 1: Direct Parent Signs Upgrade Request}
\begin{align*}
\text{Req}_{P_0} &= \{ \text{LeafHash}, \text{DesiredRole}, T, \text{Nonce} \}_{SK_{P_0}}
\end{align*}
\noindent
\textit{Only $P_0$ can sign; confers authority to upgrade the leaf.}

\paragraph{Step 2: Immediate Upper Layer Verification}
\begin{enumerate}
    \item Verify signature using $PK_{P_0}$:
    \[
        \text{if } \text{Verify}_{PK_{P_0}}(\text{Req}_{P_0}) = \text{False} \Rightarrow \text{Deny}
    \]
    \item Check that sender $P_0$ is a manager (not a leaf):
    \[
        \text{if Role}(P_0) \neq \text{Manager} \Rightarrow \text{Deny}
    \]
    \item Apply policy checks (tree depth, size, quota):
    \[
        \text{if policy violation} \Rightarrow \text{Deny}
    \]
    \item If denied, propagate \text{Deny} downward to $P_0$ and $L$; do not forward upstream.
\end{enumerate}

\paragraph{Step 3: Forward Request Upstream}
 
\begin{align*}
\text{if Deny = False:} & \\
& \text{forward } \{\text{Req}_{P_0}, \text{PolicyFlags}\} \\
& \text{to higher layers } P_2, \dots, R
\end{align*}

\paragraph{Step 4: Root Approval}
 
\begin{align*}
\text{Decision}_R = \{ &
\text{LeafHash}, \\
& \text{DesiredRole}, \\
& T, \\
& \text{Nonce}, \\
& \text{Approved/Deny} 
\}_{SK_R}
\end{align*}

\noindent
\textit{Root applies global policy checks and decides approval or denial.}

\paragraph{Step 4b: Downward Propagation of Decision}
\begin{itemize}
    \item If request was denied by any intermediate layer, the downward decision begins from that layer.
    \item Otherwise, the root propagates its decision downward through the hierarchy to $P_0$ and $L$.
\end{itemize}

\paragraph{Step 5: Direct Parent Issues Formal Upgrade Certificate}

\begin{align*}
\text{if Decision = Approved: } \quad
\text{Cert}_L = \{ &
\text{LeafHash}, \\
& \text{NewRole}, \\
& T, \\
& \text{Nonce} 
\}_{SK_{P_0}}
\end{align*}

\noindent
\textit{Leaf $L$ receives $\text{Cert}_L$ and upgrades role. If denied, no certificate is issued.}

\paragraph{Step 6: Audit Trail(optional)}
\begin{itemize}
    \item Log all requests, signature verification results, policy flags, root or intermediate approvals/denials, and certificate issuance.
\end{itemize}

\subsubsection{Protocol Flow Diagram  }

\paragraph{} 
Figure~\ref{fig:hierarchical_upgrade} illustrates the hierarchical leaf upgrade protocol. The leaf node $L$ sends a certificate request to its direct parent $P_0$ (manager). The request is then propagated upward through the hierarchy to $P_1$ and intermediate nodes $P_2 \dots P_n$, ultimately reaching the root. Each intermediate node verifies the public key and role of the requester, enforces local policy constraints, and forwards approvals  upward (if denials stop and backwards). Once the root approves the request, the decision is propagated back down through the hierarchy to the parent, which issues the certificate $\text{Cert}_L$ to the leaf. This flow ensures that leaves cannot self-promote and that all upgrades are authorized and validated at multiple layers.

 \begin{figure*}[t]  
\begin{verbatim}
Leaf L           P0 (Manager)     P1 (Immediate Upper)     P2 ... Pn       Root
   |                  |                   |                     |            |
   |                  |--Req_P0--------->|                     |            |
   |                  |                   |--Verify PK & role-->|            |
   |                  |                   |--Policy check------>|            |
   |                  |                   |                     |<--Root Approval
   |                  |<--Decision--------|<-------------------|            |
   |<--Cert_L---------|                   |                     |            |
\end{verbatim}
\caption{Hierarchical leaf upgrade flow diagram .}
\label{fig:hierarchical_upgrade}
\end{figure*}

\subsubsection{Key Principles}
\begin{enumerate}
    \item Leaf cannot self-upgrade.
    \item Direct parent ($P_0$) is the sole issuer of the upgrade certificate.
    \item Immediate upper layer verifies signature and manager role.
    \item Intermediate layers and root can veto via policy.
    \item Decision propagates downward starting from deny layer or root.
    \item Audit trail maintained at all layers.
    \item Leaf ID is protected via hash (optionally salted).
\end{enumerate}

The hierarchical upgrade protocol inherently prevents protocol violations by restricting upgrade authority to the direct parent of a leaf, while all upper layers act only as policy arbiters rather than issuers. A leaf cannot self-upgrade because any valid upgrade request must carry a signature produced by its direct parent, whose role (manager vs. leaf) is already known and verifiable by its immediate upper layer. Although higher layers and the root do not possess the public key of the issuing parent, they can still determine that the request originates from a legitimate manager—rather than a leaf—because each layer verifies the identity and role of its direct child before forwarding the request upstream. To further strengthen this guarantee and prevent forged or misattributed delegation, a zero-knowledge proof (ZKP) can be attached to the parent-signed request, allowing the parent to prove possession of a valid manager credential and parent–child relationship with the leaf without revealing its private key or full identity. This ZKP enables upper layers to validate authorization correctness even in the absence of direct key knowledge, while preserving privacy and maintaining strict hierarchical control. Consequently, any node that disobeys the protocol—by attempting self-upgrade, unauthorized signing, or role forgery—is cryptographically excluded from successful promotion.

\subsection{Action Certificate Issuance and Policy Enforcement Protocol}

This protocol governs the authorization of actions for any node (leaf or manager) through hierarchical delegation, combining parent-issued certificates, upward policy evaluation, and downward enforcement decisions.

\begin{enumerate}
    \item \textbf{Certificate Request Initiation (Node $\to$ Parent):} \\
    A node $N$ submits a certificate request containing $H(ID_N)$, the requested role or action scope, and freshness values $(T,\mathrm{Nonce})$ to its direct parent $P_0$. The node \textbf{cannot contact upper layers directly} and \textbf{cannot self-issue certificates}, because certificate validity depends entirely on the parent signature.

    \item \textbf{Local Validation by Direct Parent:} \\
    The parent $P_0$ verifies that $N$ is its immediate child and that the requested scope is within its delegated authority. If the request violates local policy, it is rejected locally and not propagated.

    \item \textbf{Certificate Proposal Construction (Parent):} \\
    If local validation succeeds, $P_0$ constructs a certificate proposal
    \[
        \mathcal{C}_0 = \{ H(ID_N), \text{Scope}, T, \mathrm{Nonce} \}
    \]
    and signs it with $SK_{P_0}$. At this stage, the certificate is not yet finalized.

    \item \textbf{Upper-Layer Endorsement (Upward Propagation):} \\
    The signed proposal $\mathcal{C}_0$ is forwarded to the immediate upper layer $P_1$. Node $P_1$ verifies that:
    \begin{enumerate}
        \item $P_0$ is its legitimate child,
        \item $P_0$ holds a manager role (not a leaf),
        \item $P_0$’s delegation rights are active.
    \end{enumerate}
    If valid, $P_1$ issues a signed endorsement
    \[
        \mathcal{E}_1 = \{  \text{Role}_{P_0}, T, \mathrm{Nonce} \}_{SK_{P_1}}
    \]
    Higher layers repeat check child's legitimate and may enforce policy strategies, such as subtree limits, risk evaluation, or global constraints.

    \item \textbf{Root or Upper-Layer Policy Decision:} \\
    Any upper layer, including the root, may issue a signed denial decision based on policy considerations, even if all certificates and endorsements are valid. If no denial occurs, an approval decision is produced at the highest required layer.

    \item \textbf{Downward Decision Propagation:} \\
    The signed approval or denial decision is propagated downward. Intermediate layers record the decision but do not modify the certificate itself.

    \item \textbf{Certificate Finalization or Abortion (Parent):} \\
    Upon receiving the decision:
    \begin{itemize}
        \item \textbf{If denied}, the parent $P_0$ aborts issuance; no certificate is delivered to $N$.
        \item \textbf{If approved}, $P_0$ finalizes the certificate by assembling
        \[
            \text{Cert}_N = \Big( \mathcal{C}_0, \{\mathcal{E}_1, \mathcal{E}_2, \dots\} \Big)_{SK_{P_0}}
        \]
        and delivers it to node $N$.
    \end{itemize}
\end{enumerate}

\textbf{Protocol Properties:}
\begin{itemize}
    \item \textbf{Universal coverage:} applies to any node, whether leaf or manager.
    \item \textbf{No self-authorization:} only direct parents issue certificates.
    \item \textbf{Issuer legitimacy:} upper-layer endorsements prove the issuing parent is authorized.
    \item \textbf{Policy sovereignty:} upper layers may refuse actions independently.
    \item \textbf{Hierarchical enforcement:} denials propagate downward; certificates propagate upward.
    \item \textbf{Misbehavior containment:} violations are detected and stopped at the earliest honest layer.
\end{itemize}

\subsection{Hierarchical Delegate-Based Certificate Validation for Node Actions}

A node $X$ receives an action request from $N$ along with a parent-issued certificate. Since $X$ cannot verify the parent directly, it forwards the request to the gateway(root can be gateway, but usaully not). The gateway then coordinates a hierarchical validation: downward propagation for local parent verification, followed by upward aggregation of approvals, and finally a gateway-signed confirmation back to $X$.

\subsubsection{Protocol Steps}

\paragraph{Step 0: Action request submission}

\[
N \rightarrow X: \text{Action}, \text{Cert}_N
\]

The certificate $\text{Cert}_N$ includes:
\begin{itemize}
    \item The hash of the node identity $H(ID_N)$.
    \item Authorized role or action scope.
    \item Freshness values: nonce and timestamp to prevent replay attacks.
    \item Endorsements from upper layers $\{\mathcal{E}_1, \mathcal{E}_2, \dots\}$ proving issuer legitimacy.
\end{itemize}

\paragraph{Step 1: Forward to gateway}

\[
X \rightarrow \text{Gateway}: \text{Action}, \text{Cert}_N
\]

\paragraph{Step 2: Downward hierarchical validation}

The gateway forwards the request downward through the hierarchy to the grandparents and parent nodes of $N$:

\[
\text{Gateway} \rightarrow P_{\text{grandparent}} \rightarrow P_0: \text{ValidateRequest}(\text{Cert}_N)
\]

 Grandparent $P_1$ checks the parent $P_0$:
    \begin{itemize}
        \item Verify that $P_0$ is a legitimate member of the tenant
        \item Check that $P_0$ has delegation rights to issue certificates
        \item Verify the integrity of $P_0$'s certificate for $N$
        \item Ensure that cert  is not self-issue certificates by $P_0$
        \item Optionally check subtree policy constraints (tree size, safety)
      \end{itemize}

  Parent $P_0$ checks the child $N$**:
    \begin{itemize}
        \item Verify that $N$’s certificate request is valid (nonce, timestamp)
        \item Check that $N$ is within the authorized role/scope     
    \end{itemize}

\paragraph{Step 3: Upward approval aggregation}

Each node checks:
\begin{itemize}
    \item That its child  is a legitimate manager/leaf
    \item Delegation rights are active
    \item Local policy compliance (tree size, safety, quotas, etc.)
    \item check the integrated $\mathcal{E}_{n-1}$,  $\mathcal{E}_n$,  $\mathcal{E}_{n+1}$  in Action certificate. Suppose this node is N. Node N knows the public keys of its parent and its children.

\end{itemize}
Each node returns a signed approval or denial to its immediate upper node:

\[
P_0 \rightarrow P_{\text{grandparent}} \rightarrow \text{Root}: \text{Approval/Denial}
\]

The root collects all upward decisions.

\paragraph{Step 4: Root downwards the decisions to gateway}

\paragraph{Step 5: Gateway issues final approval}

If all checks pass, the gateway issues a signed certificate to $X$:

\[
\begin{aligned}
\text{Cert}^{\text{Gateway}}_N &= 
\text{Sign}_{SK_\text{Gateway}}\Big( 
    H(ID_N), \\
    &\quad \text{Role}, 
    T, 
    \text{Nonce}, 
    \text{DelegationProof} 
\Big)
\end{aligned}
\]

\paragraph{Step 5: X validates and executes action}

Upon receiving the root-signed certificate, $X$ verifies:
\begin{itemize}
    \item Gateway signature validity
    \item Freshness and integrity
    \item Compliance with authorized scope and role
    \item Optional local policy constraints
\end{itemize}

If all checks succeed, $X$ executes the action; otherwise, it is rejected.

\subsubsection{Advantages of this Hierarchical Validation Protocol}

Prevents unauthorized self-issuance

Nodes cannot bypass the hierarchy or issue their own certificates.

Only the parent and higher-layer nodes can validate and approve actions.

Decouples trust from local knowledge

The receiving node X doesn’t need to know the parent’s public key.

Trust is established through hierarchical delegation proofs and final root signature.

Ensures hierarchical policy enforcement

Intermediate parents can enforce subtree-specific rules (e.g., quotas, tree depth, safety).

Policies can be applied both downward and upward, allowing dynamic control.
 
Gateway-signed certificates act as a single source of truth.

Any node can verify that an action is authorized according to the full tenant hierarchy.

Supports privacy and minimal disclosure

Node identities can remain hidden via hash commitments.

Intermediate approvals provide validity without exposing sensitive identities unnecessarily.

Resilience and auditability

All approvals/denials are logged at intermediate nodes and root.

Misbehavior (e.g., a compromised parent) can be detected through inconsistencies in approval chains.

Scalable and flexible

Can handle deep hierarchies with multiple intermediate layers.

Upward and downward flows allow both small and large trees to be validated efficiently.

\textbf{Summary:}

This protocol combines hierarchical delegation, policy checks, and gateway-signed certificates, making it secure, auditable, and privacy-preserving, while preventing leaves or lower-layer nodes from misbehaving or bypassing authorization.

\subsection{Private Tree Construction}

Each node in the PVTN independently generates its own public/private key pair.
The root node represents the top-level organizational authority and initializes
the cryptographic trust anchor for the network. Authority is delegated strictly
downward through signed delegation certificates issued by managers, and private
keys are never shared.

Public keys are treated as organizational secrets rather than globally visible
identifiers. Key visibility is restricted to direct manager–member
relationships. No external node can participate in the tree or verify membership
without explicit key disclosure by an authorized manager. This design ensures
that the PVTN remains private, non-enumerable, and resistant to external
observation or infiltration.

The private virtual tree is constructed incrementally as follows:
\begin{enumerate}
    \item The organizational root generates a public/private key pair and
    establishes the initial trust anchor of the PVTN.

    \item Each manager securely distributes its public key to prospective
    members or child nodes using trusted organizational channels (e.g.,
    authenticated internal systems or out-of-band communication).

    \item A node requests membership by submitting a join request encrypted
    with the public key of its designated parent or manager. The request
    includes the node’s public key and organizational identity attributes
    (e.g., role, department, or node type).

    \item Upon receiving the join request, the parent or manager invokes the
    join and conflict detection protocol described above. If the final decision
    is \textsf{APPROVE}, the manager issues a signed delegation certificate and
    admits the requester as a \emph{leaf node} in the PVTN. If the decision is
    \textsf{REJECT}, the join request is denied.

    \item Leaf nodes possess valid membership certificates but do not issue
    further delegations unless explicitly upgraded to managerial roles by
    their parent nodes.
    
      \item Authorized Leaf   members may be upgraded to managers using upgraded protocol; they can subsequently act as managers for subordinate nodes by distributing their public keys secretly  and issuing delegation certificates  for their members.
\end{enumerate}

\subsubsection{Membership Join}

New members can request to join a PVTN under four scenarios:
\begin{enumerate}
    \item \textbf{Member does not know the manager’s ID/IP:} The join request is propagated through the VIRGO hierarchical overlay using n-tuple nodes in upper layers, ensuring delivery while limiting network-wide broadcast.

    \item \textbf{Member knows the manager’s public IP:} The request is sent directly to the manager, minimizing latency.
    \item \textbf{Member knows the manager’s local/private IP:} The request is  propagated through the VIRGO hierarchical overlay  or gateway nodes to reach the manager securely.
    \item \textbf{Member knows the manager’s VIRGO network ID:} Using the structured overlay protocol, the member routes the request from the root along the hierarchical tree path corresponding to the manager’s ID.
\end{enumerate}

Managers verify requests and issue signed delegation certificates. Membership within a domain is largely static; revocation occurs only when a member leaves the organization voluntarily or due to termination or policy violations.

\subsubsection{Delegation Models}
PVTNs support two delegation approaches:
\begin{itemize}
    \item \textbf{Hierarchical delegation:} Members only know their immediate parent/manager. Keys and certificates above the parent remain hidden, preserving secrecy of the organizational structure and supporting high-security scenarios such as military networks.
    \item \textbf{Full-path certificates:} Delegation certificates contain the full path from root to the member, enabling independent verification without relying on the manager. This reduces bottlenecks, improves auditability, and supports distributed verification.
\end{itemize}

\subsubsection{Membership Revocation}
Membership revocation occurs under the following circumstances:
\begin{itemize}
    \item \textbf{Voluntary leave:} Members who resign or leave the organization have their certificates invalidated.
    \item \textbf{Termination or policy violation:} Managers revoke delegation certificates of members who no longer comply with organizational policies.
\end{itemize}

Revocation affects only the relevant subtree, minimizing disruption to unrelated members. Delegation certificates and key isolation ensure that revoked members cannot access PVTN resources or impersonate other participants.

\subsubsection{Overlay and Network Dynamics}
The VIRGO overlay ensures that messages, delegation certificates, and revocation notices are efficiently propagated. Cryptographic verification prevents tampering by malicious or non-compliant nodes. Overlay-level redundancy and multiple verification paths enhance resilience, while behavioral monitoring and auditing by managers help maintain operational integrity.

 \subsection{Organizational Mapping and Social-Driven Hierarchy}

In real-world society, most organizations naturally follow a hierarchical tree structure: an organization has a top-level authority, multiple levels of managers, and finally individual members at the leaves. Each manager is responsible for a subset of subordinates, and authority is delegated downward level by level.

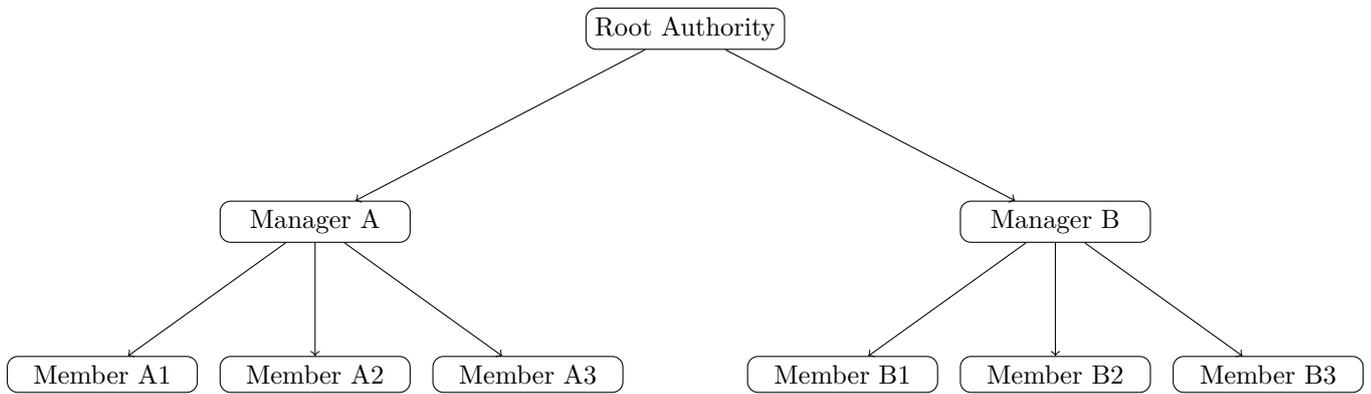
\begin{figure*}[t]
\centering
\begin{tikzpicture}[every node/.style={draw, rectangle, rounded corners, minimum width=2.5cm}, node distance=1.8cm]

\node (root) {Root Authority};

\node (m1) [below left=2cm and 2.3cm of root] {Manager A};
\node (m2) [below right=2cm and 2.3cm of root] {Manager B};

\node (c1a) [below=1.5cm of m1, xshift=-2.8cm] {Member A1};
\node (c1b) [below=1.5cm of m1, xshift=0cm] {Member A2};
\node (c1c) [below=1.5cm of m1, xshift=2.8cm] {Member A3};

\node (c2a) [below=1.5cm of m2, xshift=-2.8cm] {Member B1};
\node (c2b) [below=1.5cm of m2, xshift=0cm] {Member B2};
\node (c2c) [below=1.5cm of m2, xshift=2.8cm] {Member B3};

\draw[->] (root) -- (m1);
\draw[->] (root) -- (m2);

\draw[->] (m1) -- (c1a);
\draw[->] (m1) -- (c1b);
\draw[->] (m1) -- (c1c);

\draw[->] (m2) -- (c2a);
\draw[->] (m2) -- (c2b);
\draw[->] (m2) -- (c2c);

\end{tikzpicture}
\caption{Private Virtual Tree Network architecture with multiple  members under each manager.}
\label{fig:architecture-multiple-members-separated}
\end{figure*}

Private Virtual Tree Networks (PVTNs) as Figure~\ref{fig:architecture-multiple-members-separated}  shown are explicitly designed to reflect
\emph{social and organizational hierarchies} observed in real-world institutions.
In most organizations, authority, responsibility, and trust are delegated
top-down: from a root authority to managers, and from managers to individual
members. PVTNs directly encode this social structure as a cryptographically
verifiable tree, where each node represents an organizational entity and each
edge represents a delegated trust relationship.

A PVTN consists of:
\begin{itemize}
    \item a \textbf{root authority}, corresponding to the top-level organizational owner;
    \item \textbf{intermediate manager nodes}, representing departments, teams, or supervisors;
    \item \textbf{leaf members}, representing individual employees, services, or devices.
\end{itemize}

Each manager controls a subtree and is responsible for admitting, authorizing,
and revoking its direct subordinates. Authority is not global or flat; instead,
it follows real managerial responsibility boundaries, reducing administrative
complexity and limiting the impact of compromise.

In contrast, the underlying VIRGO network is a \emph{structural hierarchical
peer-to-peer overlay}.   VIRGO organizes nodes according to overlay identifiers
and routing efficiency, independently of social roles, administrative domains,
or trust semantics(although VIRGO can organize nodes according to social roles, here we use it scalability). Its hierarchy is designed for scalable lookup, routing, and
fault isolation rather than organizational governance.

PVTNs are therefore \emph{socially driven trust overlays} constructed on top of
the VIRGO structural overlay. The VIRGO network provides efficient hierarchical
routing and message propagation, while PVTNs impose organizational meaning,
cryptographic authorization, and privacy constraints. Multiple independent
PVTNs may coexist on the same global VIRGO overlay, enabling secure multi-tenant
and consortium deployments without revealing organizational structure or
membership relationships.

\section{ Multi-PVTN Coexistence and Isolation}
\subsection{Multi-Tenant Model}

\begin{figure*}[t]
\centering
\resizebox{\textwidth}{!}{
\begin{tikzpicture}[node distance=2cm, every node/.style={draw, rounded corners, align=center}]

\node[draw=blue!70, fill=blue!10, thick, minimum width=12cm, minimum height=6cm] (virgo) {Global VIRGO Overlay \\ (Structured Hierarchical P2P)};

\node[draw=blue!70, fill=blue!5, circle, minimum size=0.6cm] (v1) at (-4,1.5) {};
\node[draw=blue!70, fill=blue!5, circle, minimum size=0.6cm] (v2) at (-1,2.5) {};
\node[draw=blue!70, fill=blue!5, circle, minimum size=0.6cm] (v3) at (2,1.5) {};
\node[draw=blue!70, fill=blue!5, circle, minimum size=0.6cm] (v4) at (5,2.5) {};
\draw[blue!50!black, dashed, thick] (v1) -- (v2) -- (v3) -- (v4);

\node[draw=green!70, fill=green!10, thick, minimum width=3.5cm, minimum height=2.2cm, below left=2cm and -0.5cm of virgo] (pvtnA) {PVTN-A \\ Trust Domain};
\node[draw=green!70, fill=green!5, circle, minimum size=0.5cm] (a1) at (-4.2,-0.2) {};
\node[draw=green!70, fill=green!5, circle, minimum size=0.5cm] (a2) at (-3.6,-0.8) {};
\draw[green!50!black, thick] (a1) -- (a2);

\node[draw=red!70, fill=red!30, circle, minimum size=0.5cm] (a_revoked) at (-4.0,-0.5) {};
\node[draw=none] at (-4.0,-0.9) {\scriptsize Revoked};

\node[draw=orange!70, fill=orange!10, thick, minimum width=3.5cm, minimum height=2.2cm, below=2cm of virgo] (pvtnB) {PVTN-B \\ Trust Domain};
\node[draw=orange!70, fill=orange!5, circle, minimum size=0.5cm] (b1) at (-1.2,-0.2) {};
\node[draw=orange!70, fill=orange!5, circle, minimum size=0.5cm] (b2) at (-0.8,-0.8) {};
\draw[orange!50!black, thick] (b1) -- (b2);

\node[draw=red!70, fill=red!10, thick, minimum width=3.5cm, minimum height=2.2cm, below right=2cm and -0.5cm of virgo] (pvtnC) {PVTN-C \\ Trust Domain};
\node[draw=red!70, fill=red!5, circle, minimum size=0.5cm] (c1) at (2.2,-0.2) {};
\node[draw=red!70, fill=red!5, circle, minimum size=0.5cm] (c2) at (2.8,-0.8) {};
\draw[red!50!black, thick] (c1) -- (c2);

\draw[->, thick, green!50!black] (a1.north) -- ++(-0.5,1) -- (v1.south) node[midway,left]{Delegation Certificates};
\draw[->, thick, green!50!black] (a2.north) -- (v2.south);

\draw[->, thick, orange!50!black] (b1.north) -- (v2.south);
\draw[->, thick, orange!50!black] (b2.north) -- (v3.south);

\draw[->, thick, red!50!black] (c1.north) -- (v3.south);
\draw[->, thick, red!50!black] (c2.north) -- (v4.south);

\draw[green!70, dotted, thick] (a1) -- ++(-0.3,-0.5) node[draw=none,right]{Subtree Isolation};
\draw[orange!70, dotted, thick] (b1) -- ++(0.3,-0.5) node[draw=none,left]{Subtree Isolation};
\draw[red!70, dotted, thick] (c1) -- ++(-0.3,-0.5) node[draw=none,right]{Subtree Isolation};

\draw[red, ->, thick, bend left=20] (a2) to node[above]{Replay-check nonce} (v2);
\draw[red, ->, thick, bend left=25] (b2) to node[above]{Replay-check nonce} (v3);
\draw[red, ->, thick, bend left=25] (c2) to node[above]{Replay-check nonce} (v4);

\draw[->, thick, red!80!black, dashed] (a_revoked.north) -- ++(0,1) -- (v1.south) node[midway,left]{Revocation Notice};
\draw[->, thick, red!80!black, dashed] (a_revoked.north east) -- (v2.south) node[midway,right]{Revocation Notice};

\node[draw=black, fill=white, rounded corners, minimum width=5cm, minimum height=2.5cm, right=0.5cm of virgo.east] (legend) {
\textbf{Legend:} \\
\textcolor{green!70}{\rule[0.5ex]{1em}{1em}} PVTN Node / Subtree \\
\textcolor{orange!70}{\rule[0.5ex]{1em}{1em}} PVTN Node / Subtree \\
\textcolor{red!30}{\rule[0.5ex]{1em}{1em}} Revoked Node \\
\tikz[baseline=-0.5ex] \draw[red!80!black, thick, dashed] (0,0) -- (1em,0); Revocation Notice Flow \\

 \tikz[baseline=-0.5ex] \draw[blue!50!black, thick, dashed] (0,0) -- (1em,0); VIRGO Routing Links \\

 \tikz[baseline=-0.5ex] \draw[red, thick, ->] (0,0) -- (1em,0); Replay Protection Flow \\

 \tikz[baseline=-0.5ex] \draw[green!70, dotted, thick] (0,0) -- (1em,0); Subtree Key Isolation \\

};

\node[align=center, above=0.2cm of virgo.north] {\textbf{Global VIRGO Overlay}};
\node[align=center, below=0.2cm of pvtnA.south] {Secure PVTN-A communication};
\node[align=center, below=0.2cm of pvtnB.south] {Secure PVTN-B communication};
\node[align=center, below=0.2cm of pvtnC.south] {Secure PVTN-C communication};

\end{tikzpicture}
}
\caption{Multi-PVTN architecture over VIRGO overlay with color-coded legend. Revocation propagation, subtree key isolation, delegation certificates, and replay protection are illustrated. This figure clarifies trust, security, and multi-tenancy properties.}
\label{fig:multi-pvtn-virgo-legend}
\end{figure*}
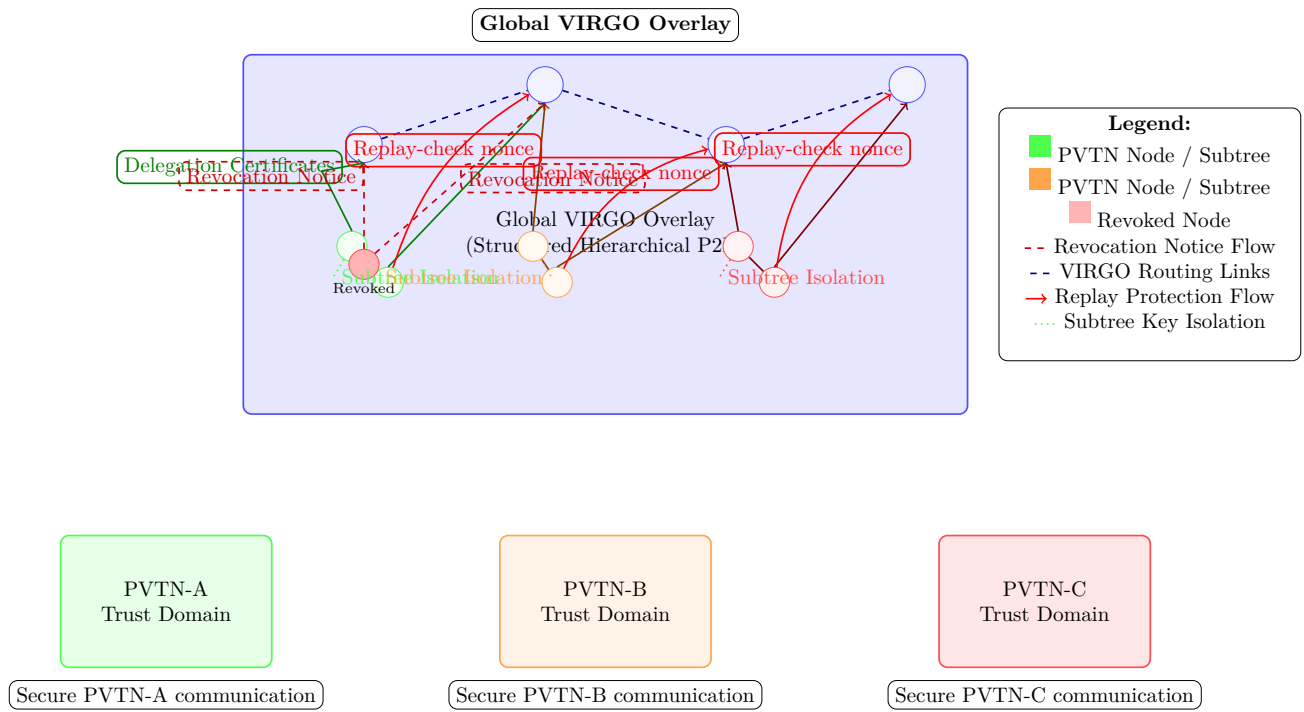

A single VIRGO overlay may simultaneously support multiple Private
Virtual Tree Networks. Each PVTN represents an independent organizational
trust domain with its own root authority, delegation hierarchy, and key
space. Although join requests and control messages traverse the same
VIRGO substrate, cryptographic isolation ensures that authorization and
membership remain strictly separated as Figure~\ref{fig:multi-pvtn-virgo-legend} shows.

The PVTN architecture operates over a shared VIRGO overlay network while
supporting multiple independent organizational tenants. Each tenant
$T_i$ is represented by an isolated \emph{Private Virtual Tree Network}
$PVTN_i$, which defines its own trust domain, membership policy, and
administrative hierarchy.

Let the root node of $PVTN_i$ be $R_i$, representing the highest authority
of tenant $T_i$. The root generates a public/private key pair
$(PK_{R_i}, SK_{R_i})$ that serves as the cryptographic trust anchor for
the entire tenant. No global root authority or system-wide certificate
authority exists.

\begin{itemize}
\item Each node $N_j \in PVTN_i$ independently generates its own key
pair $(PK_j, SK_j)$ and maintains exclusive control over its private
key.
\item Membership and authority relationships are expressed through
delegation certificates
$C_{parent \rightarrow child} =
\text{Sign}{SK{parent}}(PK_{child}, scope, constraints)$.
\item All delegation certificates are explicitly scoped to tenant
$T_i$ and are cryptographically invalid outside the issuing tenant.
\item The VIRGO overlay provides scalable routing and hierarchical
message dissemination but does \emph{not} act as a trust authority and
does not participate in cryptographic validation or authorization.
\end{itemize}

Each $PVTN_i$ therefore forms a cryptographically isolated trust tree that
shares only the routing substrate with other tenants.

\paragraph{Tenant Separation Property}
For any two tenants $T_i$ and $T_k$ with $i \neq k$, there exists no
implicit cryptographic linkage between $PVTN_i$ and $PVTN_k$. In the
absence of explicit cross-tenant delegation, certificates, public keys,
join requests, and membership information from one tenant cannot be
verified, inferred, or enumerated by nodes in another tenant.
Consequently, compromise, revocation, or structural changes within one
tenant do not affect the security or integrity of other tenants, even
though all tenants share the same VIRGO overlay infrastructure.

\paragraph{Membership Revocation Across Tenants}

Although PVTNs are designed for relatively stable organizational
membership, revocation is required to reflect real-world administrative
events such as role changes, policy violations, or credential compromise.

 Revocation Procedure

 The responsible manager removes the revoked member’s credentials and metadata from its local storage.

If the revocation is triggered by node misuse or compromise, the revocation notice is propagated to all members through the hierarchical tree.

\subsection{Cross-Tenant Authorization}

\begin{figure*}[t]
    \centering
    \includegraphics[width=\linewidth]{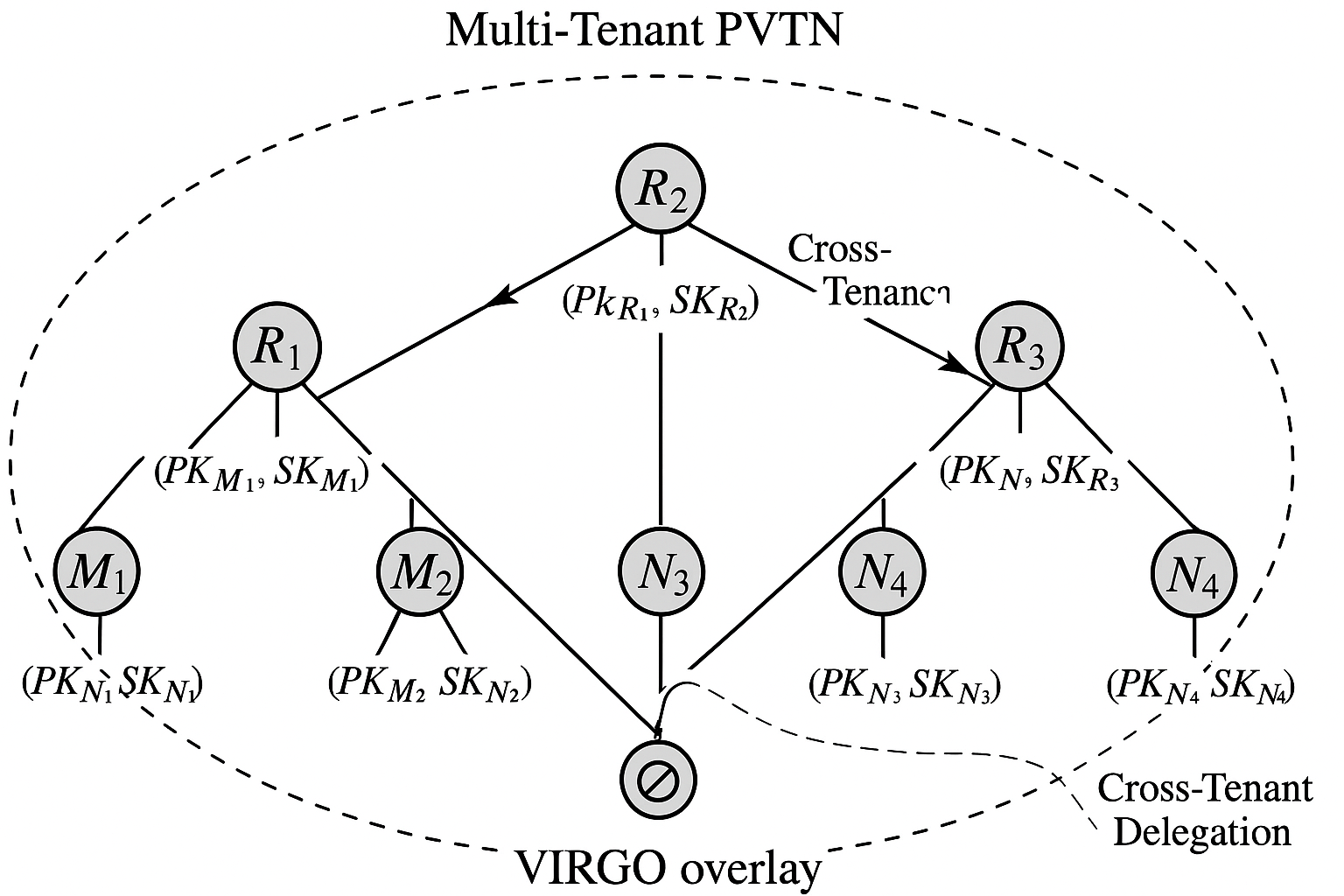}
    \caption{Cross-Tenant of  PVTNs}
    \label{fig:crosstenabtarchitecture}
\end{figure*}

While tenants are cryptographically isolated by default, PVTNs support
controlled cross-tenant collaboration through explicit authorization as Figure~\ref{fig:crosstenabtarchitecture} shows.
Such collaboration is intentional, auditable, and narrowly scoped.

\begin{itemize}
    \item Cross-tenant authorization is initiated only by tenant managers,
    typically at or near the root level.
    \item Managers exchange delegation certificates out-of-band,
    establishing limited trust relationships between specific nodes or
    roles across tenants.
    \item Delegation certificates explicitly encode scope, duration, and
    permitted actions, preventing uncontrolled trust propagation.
\end{itemize}

Nodes from tenant $T_k$ seeking access to resources in tenant $T_i$ must
present valid delegation certificates issued by an authorized manager in
$PVTN_i$. Such certificates do not merge the two trees; instead, they form
controlled trust bridges between otherwise independent hierarchies.

This approach enables inter-organizational cooperation, consortium
operation, and federated workflows without sacrificing tenant privacy,
membership confidentiality, or administrative autonomy.

While tenants are cryptographically isolated by default, PVTNs support
controlled cross-tenant collaboration through explicit authorization.

\begin{itemize}
\item Cross-tenant authorization is initiated exclusively by tenant
managers, typically at or near the root level.
\item Managers exchange delegation certificates out-of-band,
establishing narrowly scoped trust relationships between specific
nodes or roles across tenants.
\item Delegation certificates explicitly encode scope, duration, and
permitted actions, preventing uncontrolled trust propagation.
\end{itemize}

Such certificates do not merge trust trees or introduce transitive trust.
Instead, they create auditable, limited trust bridges between otherwise
independent PVTNs, preserving tenant autonomy and isolation.

\subsection{Isolation Lemmas}

 \textbf{Lemma 1 (PVTN Isolation).}
Let $PVTN_i$ and $PVTN_j$ be two distinct Private Virtual Tree Networks
operating over the same VIRGO overlay. A node authorized in $PVTN_i$
cannot obtain valid membership, delegation certificates, or topology
information of $PVTN_j$ without possession of a delegation certificate
issued by an authorized manager in $PVTN_j$.

\emph{Proof (Sketch).}
Join requests for $PVTN_j$ are encrypted using public keys known only to
authorized managers within $PVTN_j$. Delegation certificates are signed
exclusively with the corresponding managers’ private keys. Under the
Dolev--Yao adversary model, a node outside $PVTN_j$ cannot decrypt join
requests, forge certificates, or verify delegation chains without
explicit managerial authorization. Since the VIRGO overlay provides
routing only and does not participate in trust or identity validation,
access to the overlay does not reveal PVTN membership or topology.
Therefore, cryptographic isolation between $PVTN_i$ and $PVTN_j$ is
preserved. $\square$

\textbf{Lemma 2 (Cross-Tenant Join Isolation).}
Let $PVTN_i$ and $PVTN_j$ be two distinct tenants with $i \neq j$. A node
belonging to $PVTN_j$ cannot obtain a valid delegation certificate from
$PVTN_i$ without possession of a manager public key of $PVTN_i$ explicitly
disclosed through authorized delegation.

\emph{Proof (Sketch).}
 
Join requests are encrypted using the public key of an authorized manager
within the target PVTN, preventing disclosure or modification by external
nodes. Each request includes a freshly generated nonce 
r, which is cryptographically bound to the issued delegation certificate.

Upon approval, the manager generates a delegation certificate signed with
its private key, ensuring authenticity, integrity, and non-repudiation.
The manager’s response—containing the signed delegation certificate and
associated metadata—is encrypted using the requester’s public key, ensuring
that only the requesting node can access the authorization outcome.

Crucially, members of other tenants do not possess the manager’s public
key of the target PVTN. As manager public keys are treated as
organizational secrets and disclosed only within direct manager–member
relationships, nodes belonging to other tenants cannot correctly encrypt
join requests, decrypt authorization responses, or verify delegation
certificates. Consequently, cross-tenant nodes cannot obtain valid
membership certificates or infer internal membership structure.

Under the Dolev–Yao adversary model, an adversary lacking the manager’s
private key cannot forge valid delegation certificates, and an adversary
lacking either the manager’s public key or the requester’s private key
cannot participate in or observe the join protocol. The binding of the
nonce r to both the encrypted join request and the signed certificate
prevents replay of stale join messages. Together, these mechanisms ensure
confidentiality, authenticity, freshness, and strict tenant isolation.

\textbf{Lemma 3 (Loop Freedom).}
The conflict detection protocol terminates and is loop-free.

\emph{Proof (Sketch).}
Messages propagate strictly upward during aggregation and strictly
downward during dissemination, following the PVTN tree structure. No node
sends messages laterally or upward after receiving the final decision.
Since the tree is finite, the protocol terminates.

  \section{Threat Model and Security Analysis}
\subsection{Formal Threat Assumptions}

We adopt the classical Dolev–Yao adversarial model, under which the adversary has complete control over the communication network. Specifically, the adversary can eavesdrop on, intercept, delay, replay, drop, inject, and arbitrarily modify messages transmitted over the VIRGO overlay network. The adversary may also coordinate multiple malicious nodes and attempt to infer organizational structure through traffic analysis.

Cryptographic primitives are assumed to be ideal:
the adversary cannot decrypt ciphertexts without possession of the corresponding private key, cannot forge digital signatures without access to the signing key, and cannot derive private keys from public keys. Hash functions are assumed to be collision resistant.

The adversary may compromise a subset of nodes, thereby learning their private keys and all locally stored certificates. However, compromise of a node does not automatically compromise its parent, siblings, or ancestors unless explicit delegation exists. The adversary may generate arbitrarily many identities (Sybil attempts), but cannot obtain valid delegation certificates without managerial authorization.

We assume the underlying VIRGO overlay network provides reliable message routing and hierarchical addressing. Attacks targeting pure availability (e.g., flooding or routing disruption) are considered orthogonal and are outside the scope of this work.

\subsection{Security Goals}

PVTNs are designed to satisfy the following security objectives:

\begin{itemize}
\item \textbf{Confidentiality}: Membership requests and internal control messages must not be readable by unauthorized parties.
\item \textbf{Authentication}: All membership and delegation actions must be cryptographically verifiable.
\item \textbf{Authorization}: Only explicitly authorized managers can admit or revoke members.
\item \textbf{Sybil Resistance}: Identity creation alone must not grant network membership.
\item \textbf{Compromise Containment}: Breaches must be limited to the affected subtree.
\item \textbf{Non-enumerability}: External observers must not be able to enumerate members or organizational structure.
\end{itemize}

\subsection{Security Analysis}

\paragraph{Confidentiality of Join Requests}
Join requests are encrypted using the designated manager’s public key. Since public keys are not globally visible and are only disclosed within direct manager–member relationships, unauthorized nodes cannot decrypt join requests or infer organizational membership. This protects against passive eavesdropping and traffic inspection attacks.

\paragraph{Authentication and Authorization}
Authorization is enforced through manager-signed delegation certificates. Each certificate cryptographically binds a child’s public key to a parent’s authority. Verification requires a valid signature chain rooted at a trusted manager, preventing unauthorized membership even if network routing is compromised.

\paragraph{Replay Protection}
Each join request includes a cryptographically random nonce or timestamp bound to the request contents. Managers maintain a local cache of recently seen nonces to prevent replay attacks. Replayed messages are rejected without affecting existing memberships.

\paragraph{Sybil Resistance}
While an adversary may generate an arbitrary number of public/private key pairs,
identities alone convey no trust in PVTNs. Membership is granted only through the
explicit issuance of a delegation certificate by an authorized manager, reflecting
human or organizational approval. Consequently, Sybil identities cannot infiltrate
the PVTN without repeatedly obtaining valid managerial authorization. The absence
of a global identity namespace and the use of manager-scoped delegation further
prevent Sybil nodes from accumulating influence or bypassing access control.

\paragraph{Compromise Containment}
If a node is compromised, the adversary gains access only to that node’s private key and certificates. Because delegation is strictly hierarchical, the compromise affects only the node’s subtree. Ancestors and sibling subtrees remain unaffected, preserving global security.

\paragraph{Non-enumerability and Privacy}
Public keys and certificates are not published in global directories. External observers cannot enumerate nodes or infer tree structure, even with full network visibility. This property is critical for privacy-sensitive environments such as enterprises, research consortia, and military networks.

\subsection{Threat Model Summary}

\begin{table*}[t]
\centering
\caption{Threat Model and Mitigations in Private Virtual Tree Networks (PVTNs)}
\begin{tabular}{|l|p{6.2cm}|p{5.6cm}|}
\hline
\textbf{Threat} & \textbf{Description} & \textbf{Mitigation} \\
\hline
Eavesdropping
& Interception of join, delegation, or control messages over the public VIRGO overlay
& Join requests encrypted with manager public keys; no plaintext exposure \\
\hline
Replay Attack
& Reuse of previously valid join or authorization messages to gain unauthorized access
& Nonce or timestamp binding; replay detection at manager nodes \\
\hline
Forgery
& Creation of fake membership, delegation, or revocation certificates
& Delegation certificates signed with manager private keys; signature verification \\
\hline
Sybil Attack
& Generation of multiple fake identities to gain influence or unauthorized access
& Manager-controlled authorization; identities require valid delegation certificates \\
\hline
Node Compromise
& Exposure of a node’s private key due to compromise or insider threat
& Cryptographic isolation at subtree level; no implicit trust propagation \\
\hline
Key Exposure
& Long-term leakage of cryptographic keys through operational or cryptanalytic means
& Periodic key rotation and certificate expiration within subtrees \\
\hline
Unauthorized Cross-Tenant Access
& Attempts by nodes from other tenants to join or enumerate a PVTN
& Explicit cross-tenant delegation only; default cryptographic isolation \\
\hline
Malicious Overlay Behavior
& Non-compliant or malicious VIRGO nodes attempting message manipulation
& End-to-end cryptographic verification independent of overlay routing \\
\hline
\end{tabular}
\label{tab:threat-model}
\end{table*}

The table 1 summarizes how PVTNs enforce confidentiality, integrity, and authorization
despite operating over a fully adversarial network. All security guarantees are
cryptographically enforced at the application layer and do not rely on the correctness
or trustworthiness of the underlying VIRGO overlay.

\subsection{Periodic Key Rotation}

To mitigate long-term key exposure, PVTNs support localized periodic key rotation. Each manager periodically generates a fresh public/private key pair and securely distributes the new public key to its direct subordinates. New delegation certificates are issued under the updated key, while older certificates naturally expire.

Key rotation is strictly local to the affected subtree and does not require global coordination or rekeying across tenants. This design minimizes operational disruption while significantly reducing the attack window associated with compromised long-term keys.

\subsection{Security Guarantees}

Under the stated assumptions, PVTNs provide:

\begin{itemize}
\item Cryptographically enforced hierarchical authorization
\item Strong privacy through non-enumerable membership
\item Robust resistance to Sybil and forgery attacks
\item Damage containment under partial compromise
\end{itemize}

These properties collectively demonstrate that PVTNs offer a secure and scalable foundation for multi-tenant, hierarchy-driven distributed systems operating over public infrastructures.
 
\section{Deployment Scenarios}

Consider a large enterprise or a multi-institution research consortium operating over a shared public infrastructure. The VIRGO overlay provides scalable routing and hierarchical organization across all participants. Each department, subsidiary, or partner organization establishes its own Private Virtual Tree Network (PVTN), rooted at its respective authority or manager.

In an enterprise setting, employees join by submitting encrypted requests through the VIRGO overlay. Only designated managers can decrypt and approve membership, issuing delegation certificates scoped to organizational roles. Departments remain cryptographically isolated, ensuring that internal organizational structures and membership remain private from external observers.

In a research consortium, each institution operates its own PVTN over the shared VIRGO substrate. Cross-consortium collaboration is enabled through controlled delegation between managers without exposing global membership or relying on a central authority. When a member leaves or a credential is compromised, revocation affects only the relevant subtree, minimizing operational disruption.

This deployment model supports secure collaboration, delegated administration, and privacy-preserving multi-tenancy, all without reliance on global certificate authorities or centralized identity providers.

 \subsection{Gateway-Mediated Hierarchical Verification for Storage Access}

In this scenario, a child node presents a certificate issued by a manager to a storage node. The storage knows only the gateway's public key. To verify the certificate, the gateway coordinates hierarchical validation: the request is propagated downward to grandparents and parents, who check certificate validity and delegation, and then approvals propagate back up to the gateway. Both child and manager identities remain hidden to preserve privacy.

\subsubsection{Protocol Steps}

\paragraph{Step 0: Child submits request}
\[
\text{Child} \rightarrow \text{Storage}: \text{Cert}_{m\rightarrow c}
\]

The certificate includes:
\begin{itemize}
    \item A cryptographic commitment to the child identity $\text{Commit(child\_id)}$.
    \item Permissions for storage access.
    \item Nonce and validity period to prevent replay.
     \item Endorsements from upper layers $\{\mathcal{E}_1, \mathcal{E}_2, \dots\}$ proving issuer legitimacy.
     \item Zero-knowledge proof $\text{ZKP}_{\text{manager}}$ showing the manager belongs to the tenant.
\end{itemize}

\paragraph{Step 1: Storage forwards certificate to gateway}
\[
\text{Storage} \rightarrow \text{Gateway}: \text{Cert}_{m\rightarrow c}, \text{Storage\_ID}, \text{context}
\]

\paragraph{Step 2: Gateway upward to hierarchy}
The gateway forwards the request upwards to the root if needed (or coordinates directly with upper layers) and initiates downward verification to grandparents and parents:

\[
\text{Gateway} \rightarrow P_{\text{grandparent}} \rightarrow P_0: \text{ValidateRequest}(\text{Cert}_{m\rightarrow c})
\]

\paragraph{Step 3: Grandparent and parent verification}

Each intermediate node performs checks:

\begin{itemize}
    \item \textbf{Grandparent $P_1$}:
    \begin{itemize}
        \item Verify $P_0$ is a legitimate member of the tenant.
        \item Check delegation rights of $P_0$ to issue certificates.
         \item Ensure that cert is not   self-issue certificates by $P_0$.

        \item Optionally enforce policy constraints (subtree size, safety).
    \end{itemize}
    \item \textbf{Parent $P_0$}:
    \begin{itemize}
        \item Verify the child $N$’s certificate for integrity and freshness (nonce, timestamp).
        \item Check that $N$ is authorized for the requested action.
        \item generate a random number , forward with valid flag to gateway.
         \item  in the same time , sign the random number with public key of N , and send to N.
        
            \end{itemize}
\end{itemize}

\paragraph{Step 4: Approval propagation upward}

Validations propagate upward to the gateway as approvals or denials:

\[
P_0 \rightarrow P_{\text{grandparent}} \rightarrow \text{Gateway}: \text{Approval / Denial}
\]

\paragraph{Step 5: Gateway issues signed proof for storage}

The gateway generates a signed proof that storage can verify using its public key:

\[
\begin{aligned}
\text{SignedProof} &= 
\text{Sign}_{SK_\text{Gateway}}\Big(
    H(\text{Cert}_{m\rightarrow c}), \\
    &\quad \text{DelegationProof}, \\
    &\quad \text{Storage\_ID}, \\
    &\quad \text{Nonce}, \\
    &\quad \text{ValidityPeriod},\\
     &\quad \text{Random Number by  $P_0$ }\\
\Big)
\end{aligned}
\]

\paragraph{Step 6: Storage verification and access grant}

Storage verifies the gateway signature. If valid, 
It asks N to send the random number received.  Then it compares the number from N and the number in cert from gateway.  If same, then proof  N is real by using its own private key to decrypt message to get the number.    and is in the same Tenant. Then
it grants access according to the approved permissions. Storage does not need to know the public keys of any intermediate nodes.

\subsubsection{Security and Privacy Properties}

\begin{itemize}
    \item \textbf{Child identity privacy}: Cryptographic commitment in the certificate.
    \item \textbf{Manager identity privacy}: Zero-knowledge proof in the certificate; opaque delegation proofs.
    \item \textbf{Hierarchical validation}: Grandparent and parent nodes validate certificate and delegation before approval.
    \item \textbf{Gateway trust anchor}: Storage trusts only the gateway’s signature.
    \item \textbf{Proof authenticity}: Manager signs certificate hash; parents/grandparents approve validity.
    \item \textbf{Replay protection}: Nonce and validity period prevent certificate reuse.
    \item \textbf{Access control}: Storage grants access solely based on gateway-verified proof.
\end{itemize}
  
\subsubsection{Usefulness and Practical Benefits}

The gateway-mediated hierarchical verification protocol provides significant advantages in distributed systems. Storage nodes do not need to know the public keys of intermediate nodes; they rely solely on the gateway's signature, which greatly simplifies key management in large hierarchies. Delegation rules, subtree policies, and safety constraints are enforced by grandparents and parents, ensuring that only authorized nodes can issue certificates or perform actions. Privacy is preserved through zero-knowledge proofs and cryptographic commitments, hiding the identities of both children and managers while still enabling secure verification. 

The protocol mitigates misbehavior by preventing leaves or compromised parents from bypassing hierarchical checks, as approvals must propagate through multiple layers before a gateway-issued certificate is trusted. It is scalable because initial certificate validation is delegated to intermediate nodes, and broadcasts are limited to relevant branches, making it suitable for large-scale systems. Upper layers can flexibly enforce policies such as maximum tree depth, subtree size limits, temporal restrictions, or risk-based constraints without exposing internal node identities to storage. Additionally, storage nodes can execute actions solely based on the gateway-signed certificate, allowing lightweight operation without maintaining complex trust relationships. Hierarchical delegation proofs also provide auditability, enabling accountability while maintaining privacy.

This approach is particularly useful for multi-tenant cloud storage, where access policies must be enforced without revealing the internal structure or member identities of a tenant. It is also well-suited for IoT networks with lightweight devices that cannot store all public keys of hierarchical managers, as well as collaborative platforms where enterprises with hierarchical structures need to delegate permissions securely. Finally, the protocol ensures secure content distribution, as certificates issued through hierarchical delegation guarantee that only authorized nodes can access or modify content while minimizing trust exposure.

\section{Discussion}
By aligning cryptographic trust with managerial authority, PVTNs reflect real operational practices and reduce administrative complexity while preserving scalability. Unlike traditional PKI-based or flat trust models, PVTNs restrict key visibility and authorization to direct manager–member relationships, preventing global enumeration of participants and mitigating the risk of unauthorized access.

The hierarchical structure of PVTNs, supported by the VIRGO overlay, enables efficient routing, delegation, and dynamic membership management. Subtree-based revocation ensures that security incidents, such as compromised credentials or departing members, affect only the relevant portion of the network, minimizing operational disruption. This approach also facilitates fine-grained access control within multi-tenant or consortium environments without relying on centralized authorities.

Moreover, PVTNs provide a flexible framework for cross-organizational collaboration. Controlled delegation between managers allows secure, auditable interactions between separate virtual trees while preserving privacy and isolation. By combining overlay-based scalability with cryptographic enforcement, PVTNs offer a practical solution for secure, dynamic, and privacy-preserving distributed systems, bridging the gap between theoretical trust models and real-world hierarchical organizations.
  
The model assumes that the VIRGO overlay operates reliably and that participating nodes follow protocol specifications. Malicious or non-compliant nodes may attempt to disrupt routing or authorization processes. To mitigate such risks, PVTNs rely on cryptographic verification of all membership operations and delegation certificates, allowing invalid or tampered messages to be detected and discarded. In addition, overlay-level redundancy and multiple independent verification paths provided by VIRGO improve tolerance to node failures and misbehavior. Behavioral monitoring and auditing by managers or designated supervisory nodes can further identify protocol violations and trigger the isolation or revocation of misbehaving participants.

  \subsection{Limitations}
While PVTNs provide strong security guarantees and scalable management, several limitations remain.

First, cryptographic operations associated with new member joins introduce additional computational and communication overhead. Although join events are typically infrequent, the cost may become noticeable in highly dynamic environments. When a joining node does not know the identifier or network location of the responsible manager, the join request is routed through the VIRGO hierarchical overlay using $n$-tuple nodes in upper layers. This hierarchical propagation ensures that the request eventually reaches the appropriate manager without resorting to network-wide broadcast; however, it may increase message size and join latency, particularly in large-scale deployments.

Second,  there are  malware or virus-infected nodes, and delayed detection of subtle protocol deviations, particularly in highly dynamic or large-scale environments. While cryptographic verification can prevent unauthorized actions, compromised nodes may still behave correctly at the protocol level while attempting to disrupt availability or leak information through side channels.

Addressing such threats may require stronger adversarial models, enhanced behavioral monitoring, intrusion or anomaly detection mechanisms, and additional fault-tolerance techniques. These extensions, along with defenses against large-scale coordinated malware propagation, are left for future work.

 \section{Conclusion}
PVTNs leverage the VIRGO overlay network to construct private, cryptographically enforced hierarchical trust domains over open and untrusted infrastructures. By encrypting join requests using manager public keys and enforcing authorization through signed delegation certificates, PVTNs provide fine-grained, verifiable control over membership and access while preserving scalability. In addition, the design supports public key hiding, ensuring that node public key is disclosed only to authorized parent and its children , thereby reducing information leakage and limiting adversarial reconnaissance.

Authorization decisions are governed by action certificates that explicitly bind permitted operations to delegated identities. Each action certificate is cryptographically verifiable and validated along the delegation chain before execution, ensuring that all actions conform to current authorization policies. Certificate validity checks, including signature verification, freshness, and revocation status, are enforced at each decision point, preventing the misuse of stale or forged credentials.

The proposed architecture integrates the scalability and routing efficiency of VIRGO with strong security guarantees derived from public-key cryptography and explicit delegation chains. Unlike flat PKI-based or fully decentralized trust models, PVTNs achieve a balance between administrative control and distributed operation, making them suitable for enterprise, consortium, and multi-tenant environments that require both autonomy and accountability.

Importantly, PVTNs are not visible to non-member nodes. Membership information, public keys, and internal topology are disclosed only to authorized participants, preventing external entities from discovering or enumerating private virtual trees. Moreover, PVTNs provide a flexible framework for cross-organizational collaboration: controlled delegation between managers enables secure and auditable interactions across separate virtual trees while preserving privacy and isolation. By combining overlay-based scalability with cryptographic enforcement, PVTNs offer a practical solution for secure, dynamic, and privacy-preserving distributed systems, effectively bridging the gap between theoretical trust models and real-world hierarchical organizations.

Future work will focus on a comprehensive performance evaluation of PVTNs under realistic workloads, including join latency, latency of certificate   issued and   check ,  and cryptographic overhead. Optimizing cryptographic operations and message aggregation for large-scale deployments remains an important direction.  Additionally, integrating PVTNs with existing middleware and cross-tree federation mechanisms will expand applicability to multi-organization ecosystems. Research into resilience against active adversaries and potential extensions to support attribute-based or policy-driven access control is also planned.
 
\section {Declaration of generative AI and AI-assisted technologies in the manuscript preparation process}
Statement: During the preparation of this work the authors used ChatGPT   in order to prepare and writing the draft paper. After using this tool, the authors reviewed and edited the content as needed and take full responsibility for the content of the published article.

\end{document}